# Biogeography-Based Optimization and Support Vector Regression for Freeway Travel Time Prediction and Feature Selection

**Prateek Bansal**
Graduate Student
Department of Civil, Architectural and Environmental Engineering
The University of Texas at Austin, Texas, United States
Tel: +1-512-293-1802
Email: prateekbansal@utexas.edu

**Last edited:** September 14, 2013

**ABSTRACT**
As travelers make their choices based on travel time, its prior information can be helpful for them in making more informed travel decisions. To achieve this goal, travel time prediction models have been proposed in literature, but identification of important predictors has not received much attention. Identification of important predictors reduces dimensions of input data, which not only lessens computational load, but also provides better understanding of underlying relationship between important predictors and travel time. Moreover, collection of only important predictors can lead to a significant equipment savings in data collection. Therefore, this study proposes a hybrid approach for feature selection (identifying important predictors) along with developing a robust freeway travel time prediction model. A framework integrating biogeography-based optimization (BBO) and support vector regression (SVR) has been developed. It was validated by predicting travel time at 36.1 km long segment of National Taiwan Freeway No. 1. The proposed hybrid approach is able to develop a prediction model with only six predictors, which is found to have accuracy equivalent to a stand-alone SVR prediction model developed with all forty three predictors.

# 1 INTRODUCTION

There is an increasing need for advanced transportation management and information systems that can provide travelers and traffic managers with accurate and reliable real-time traffic information (1). According to the underlying assumptions of most route choice models, road users are rational decision makers and make their choices based on minimizing expected costs in terms of travel time. Thus, providing travelers with travel time information allows them to make more informed decisions, yielding not only cost-benefits for the individual, but also more stable and less congested traffic conditions. Several researchers have been trying to get best travel time prediction methods from past few decades, to name a few- Kalman filtering (2 and 3), locally-weighted regression (4), Box–Jenkins time series (5 and 6), autoregressive integrated moving average (7), Markov chains (8), and data fusion algorithm (9).

The prediction of long-distance travel time is a major concern because longer freeway sections contain more interchanges, which lead to more complex changes in traffic flow characteristics. Moreover, longer durations of forecasting periods may encounter 'non-recurrent congestion' as a result of unprecedented incidents or accidents. Diurnal non-recurrent congestion events on freeways make it more challenging to develop robust travel time prediction model. Previous studies suggest that neural networks (10 and 11) and support vector regression (SVR) (12) preserve a high degree of robustness and prediction accuracy in such complex and nonlinear scenarios.

Along with development of accurate and robust travel time prediction models, identification of important predictors is equally important. Identification of important predictors or elimination of 'noisy' predictors reduces the dimensions of input data, which not only moderates computational load, but also provides clearer understanding of underlying relationships between important predictors and target variable. It can potentially allow researchers to collect only the more relevant variables in the future which can lead to significant equipment savings in data collection. Moreover, data collected from automated vehicle identification (AVI) systems have issues of missing data. If one can identify locations of important devices, he or she can develop a robust imputation method to retrieve the missing data. Additionally, higher order precision can be achieved during collection of data corresponding to the important predictors.

Several feature selection algorithms have been developed and employed in many areas of engineering (13 and 14), but relatively very few studies (15) explored it in context of travel time prediction. Therefore, this study proposes a hybrid approach for feature selection (identifying important predictors) while developing an accurate and robust long-distance freeway travel time prediction model under non-recurrent congestion conditions. A framework integrating two state-of-art algorithms, biogeography-based optimization (BBO) (16) and SVR, has been developed and applied to predict travel time at the 36.1 km long section of National Taiwan Freeway No. 1.

Simon (16) developed BBO algorithm in 2008 and compared the performance of BBO against several other optimization techniques such as ant colony optimization, differential evolution, evolutionary strategy, genetic algorithm, probability based incremental learning, particle swarm optimization, and stud genetic algorithm using fourteen benchmark functions. BBO performed the best on seven out of fourteen benchmarks, assuring its competitiveness with other algorithms. Since then, it has been applied extensively in various combinatorial optimization problems such as satellite image classification (17), economic load dispatch problem (18), gene selection (19) and so forth. However, to the author's best knowledge, BBO has not been employed in any of earlier transportation related research and this serves as a supplementary motivation to explore its potential applications in advanced traveler information systems (ATIS).

The organization of remaining paper is as follows: Section 2 focuses on data collection, missing data imputation, and travel time computation; Section 3 gives theoretical background; Section 4 describes the proposed method; Section 5 implements the proposed approach on study area and compares the results of numerical experiments; and Section 6 draws conclusions illustrating the practical implications of this research.

## 2 DATA COLLECTION AND SUMMARIZATION

### 2.1 Study Area Description

In this study, National Taiwan Freeway No. 1 was chosen for data collection. It is 373 km (232 mi) long and has 20 toll stations. The selected freeway segment includes six interchanges and two system interchanges with total length of 36.1 km as shown in Figures 1 and 2. The chosen freeway segment is the busiest one, accounting for around 23.5% of the average daily traffic volume. The data were collected from September 16 to October 16, 2009 between the Yangmei and Taishan toll stations in north direction (Figure 1). The databases used in this study are established by Taiwan's governmental agencies for information dissemination, and research use.

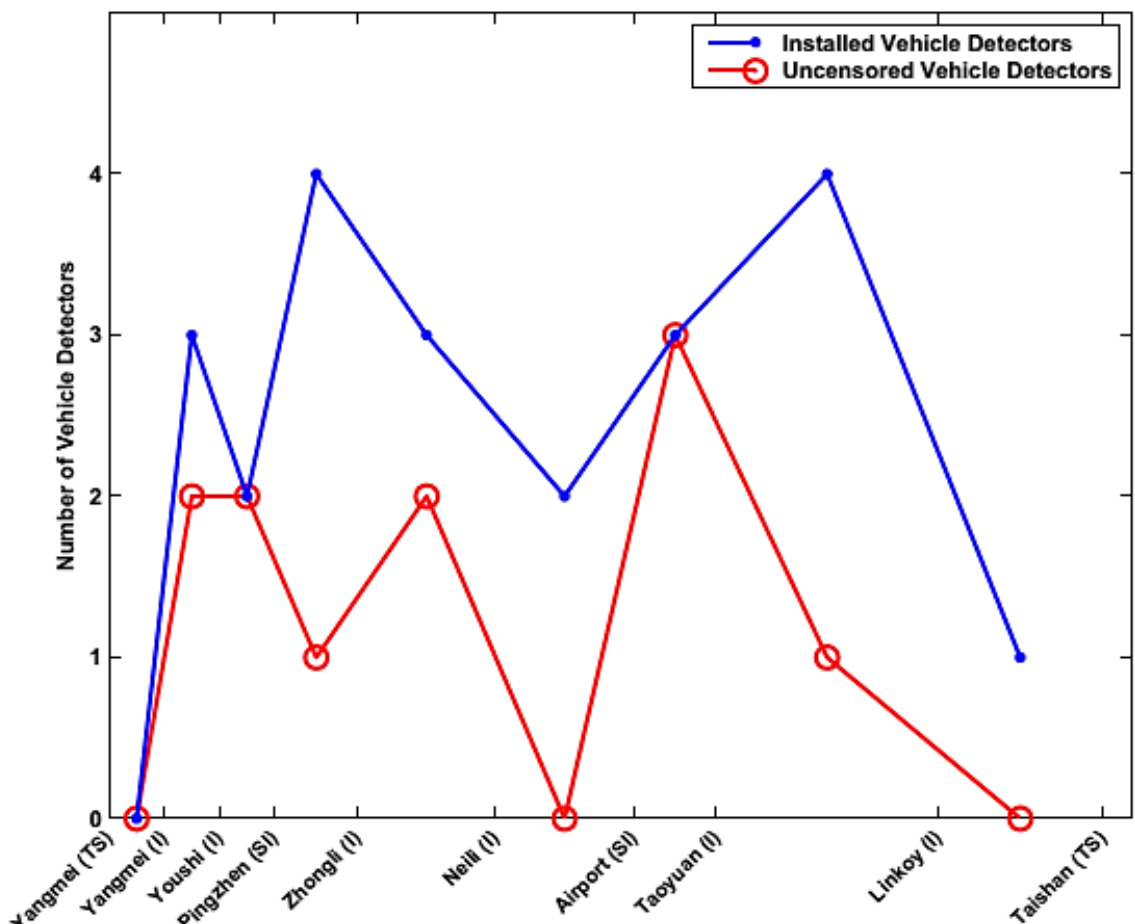

**FIGURE 1 Selected freeway section layout and number of vehicle detectors at the important locations (TS: Toll Station, SI: System Interchange, I: Interchange)**

## 2.2. Data Collection

The speed and heavy vehicle volume (flow) data were collected at a 5-minute interval by 22 dual-loop vehicle detectors (VDs) through the database of Traffic Control Center of Taiwan Area National Freeway Bureau, MOTC. In such complex scenarios, stability of VDs plays an important role and therefore, VDs having missing data for duration of more than 2 hours were eliminated. The detailed description of missing data imputation algorithm can be found in Section 2.3. After the elimination of VDs, 11 VDs were left and their relative positioning (based on distances among them) is shown in Figure 1. Among these 11 VDs, the last VD was unstable with respect to flow measures and therefore, only speed data was extracted from it, i.e. 11 speed and 10 flow variables were extracted from VDs. The electronic toll collection (ETC) charging time of freeway users was collected to calculate the historical travel time (HTT) and actual travel time (ATT). Section 2.4 describes computation of HTT and ATT from ETC charging data.

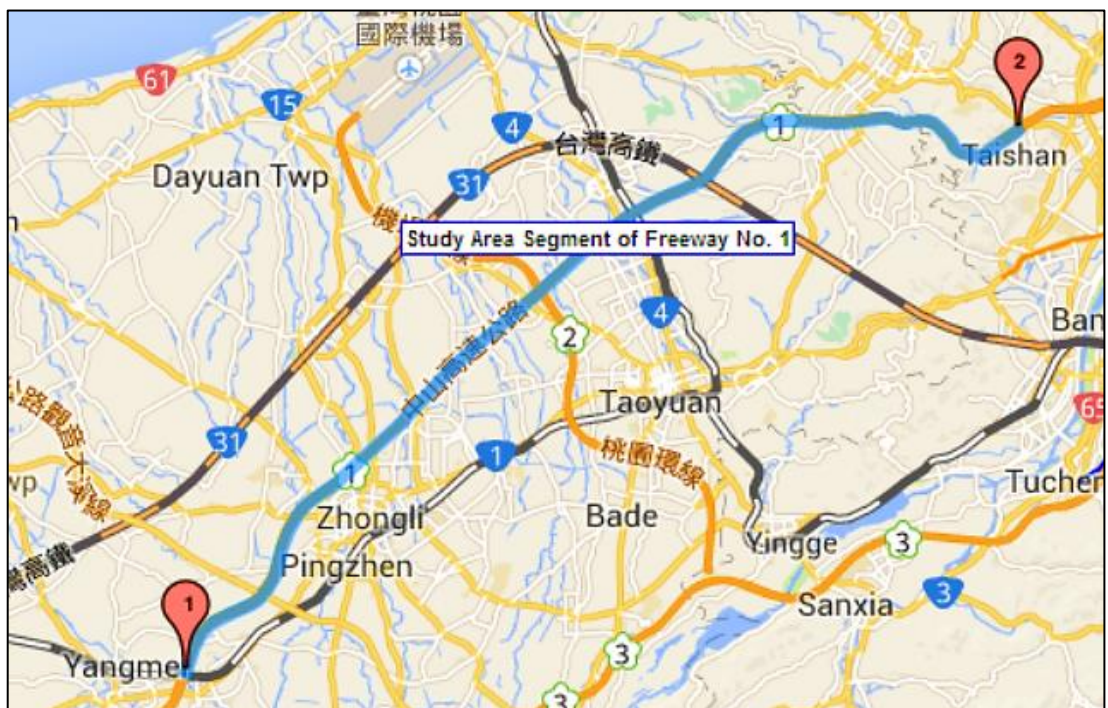

**FIGURE 2 The selected freeway segment of National Taiwan Freeway No. 1**

Taiwan's climate has frequent occurrence of rains and thus, rainfall can be an important factor in travel time prediction. Therefore, rainfall data was collected at three interchanges, viz. Yangmei, Taoyuan, and Linkou. Generally, the effects of rainfall manifest after a certain duration and therefore, rainfall data was converted into 7 cumulative rainfall variables (5 minute, 1 hour, 2 hours, 3 hours, 4 hours, 5 hours, and 6 hours) for each of the three interchanges.

There were a total of 76 accidents in the time span of data collection, and 176 vehicles were damaged and 6 people injured. Since data of relevant variables (e.g., the accident occurrence time, number of closed lanes, accident removal time etc.) for estimating the accident impact on traffic flow cannot be obtained accurately, this study utilized the speed and flow data to represent traffic characteristics, reflecting the effects of accident occurrence. The final dataset had 7908 observations and 43 variables (see Table 1) for developing the travel time prediction model.

**TABLE 1 Description of variables involved in prediction model**

| Variable | Source | Number | Input(I)/Predicted(P) |
|----------|--------|--------|-----------------------|
| Speed | Vehicle Detectors | 11 | I |
| Flow | Vehicle Detectors | 10 | I |
| Rainfall | Rainfall Detector | 21 | I |
| Historical Travel Time (HTT) | ETC System | 1 | I |
| Actual Travel Time (ATT) | ETC System | 1 | P |

## 2.3 Missing Data Imputation

Although VDs facilitated automatic data collection, but serious missing data issues occurred due to the events like poor weather conditions, maintenance, or cable thefts. Missing data of VDs can be retrieved using following three possible datasets: a) historical data of the same time on different days of the original VD (facing missing data problem); b) available data of previous time steps of the original VD using methods like arithmetic mean; c) data collected from closer upstream and downstream VDs.

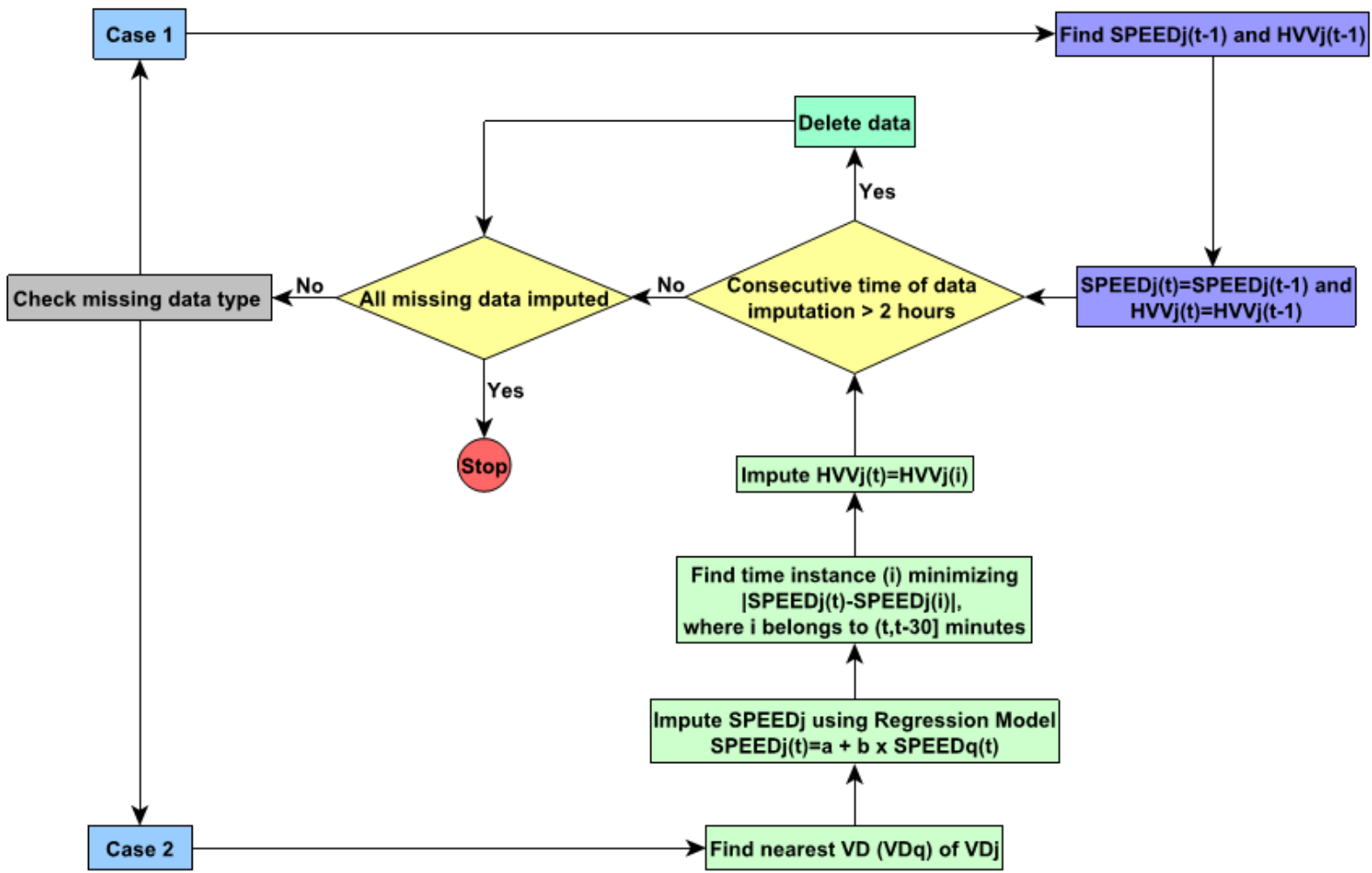

**FIGURE 3 Missing data imputation algorithm**
(Note: *HVVj(t)* is heavy vehicle volume of $j^{th}$ VD at time *t* and similarly, SPEEDj(t) is the speed detected by $j^{th}$ VD at time t)

If ATT and HTT data (collected from ETC) at time $t$ are missing then VD data of that time period is removed. This strategy was adopted to maintain consistency as ATT and HTT are the predictor and target variables in the model, respectively. Further, the missing data of VDs can be characterized in following two ways: **Case 1:** Vehicle detector $j$ ($VD_j$) has a single missing data at time $t$, but data at time *t-1* is available; **Case 2:** $VD_j$ has many missing data instances, and at least one VD does not have missing data at those instances.

Missing speed and flow data of case 2 is imputed using data of nearest VD. To impute missing flow (or heavy vehicle volume) data of $VD_j$ at time $t$, the flow at the time with the speed closest to that of time $t$ within the previous half hour is used, i.e. if the speed at time *t-1* is closest to the speed at time $t$, the missing flow at time $t$ is imputed by flow at time *t-1*. This method allowed us to control the factors such as disparity in detection quality of VDs and different traffic characteristics. The complete data imputation algorithm can be found in Figure 3. If the missing data of VD does not fall into any of two cases then it is deleted. Moreover, the VD data points with abnormally high driving speed (greater than speed limit i.e., 120 Km/hour) are also deleted.

### 2.4 Historical and Actual Travel Time Computation

To compute HTT and ATT from Yangmei (Point A) to Linkou (Point B) toll stations using collected ETC charging times, Transmit algorithm with an interval of five minutes was adopted (28). The data of vehicles passing through downstream point B in an interval of 5 minutes are collected and average travel time of the completed trips between upstream point A and downstream point B is utilized as the historical travel time of the interval. It is formally written in Equations 1 and 2:

$$HTT_{ABsi} = \{t_{Bi} - t_{Ai} | \ t - t_r \leq t_{Bi} \leq t \ and \ .6 \times HTT_{AB(s-1)i} \leq t_{Bi} - t_{Ai} \leq 1.4 \times HTT_{AB(s-1)i}\} \quad (1)$$

$$HTT_{ABs} = \left\{ \frac{\Sigma_i HTT_{ABsi}}{n_s} \right\} \quad (2)$$

where $HTT_{ABsi}$ is a historical travel time of vehicle $i$ between point A and B in $s^{th}$ time interval, $t_{Bi}$ is the time of vehicle $i$ passing through point B, $t_{Ai}$ is the time of vehicle $i$ passing through point A, $t_r$ is an observation window or length of time intervals, which is 5 minutes in this study, $HTT_{ABs}$ is an average historical travel time of $s^{th}$ time interval, $n_s$ is the total samples (vehicles completing trip) in the $s^{th}$ time interval.

The second condition in Equation 1 implies that if the travel time of the vehicle is more or less than the 40% of average HTT of the last time interval, sample point will be eliminated considering it an abnormal behavior. In other words, if the average HTT of the last time interval is 25 minutes, a trip with the travel time of more than 35 or less than 15 minutes will be eliminated.

Similarly, the data of vehicles passing through upstream point A in an interval of 5 minutes are collected, and average travel time of the completed trips between upstream point A and downstream point B are utilized as the actual travel time of the interval. It is formally written in Equations 3 and 4.

$$ATT_{ABsi} = \{t_{Bi} - t_{Ai} | \ t - t_r \leq t_{Ai} \leq t \ and \ .6 \times ATT_{AB(s-1)i} \leq t_{Bi} - t_{Ai} \leq 1.4 \times ATT_{AB(s-1)i}\} \quad (3)$$

$$ATT_{ABs} = \left\{ \frac{\Sigma_i ATT_{ABsi}}{n_s} \right\} \quad (4)$$

# 3 THEORETICAL BACKGROUND

Assuming that reader has preliminary background in machine learning, this section only covers the key theoretical concepts and formulations of BBO in the context of this study.

## 3.1 Biogeography-based Optimization (BBO)

BBO is a global optimization algorithm based on biogeography theory (16) and is used in the study for feature selection. It is based on the idea of probabilistic sharing of information between candidate solutions based on their fitness values. In this approach, each solution is described as a habitat and every habitat is geographically isolated with other habitats. The habitat suitability index (HSI) indicates the residential conditions of habitat for biological species. The features (speed, flow, and rainfall etc.) responsible for HSI of the habitat are called suitability index variables (SIVs). HSI is a function of SIV and this functional relationship dependent on the specific application.

$$HSI_i = f(\boldsymbol{SIV_i}) \tag{5}$$

These habitats are allowed to evolve generation-by-generation. The natural activities of habitats are replicated by operations like migration, mutation, and elitism, which are described in the next subsection. In this study, the objective of the BBO algorithm is to improve suitability of habitat (HSI) over a number of generations and find SIVs (important features) corresponding to the habitat having the highest HSI in the last generation (see Section 4.2 for details).

### 3.1.1 Migration

The improvement in HSI after each generation is possible by operations like immigration and emigrations of species among habitats, replicating natural form of geographical migration. Naturally, the habitats with high HSI are likely to have more species, low immigration rate, high emigration rate (due to virtue of saturation with species) and vice versa. The immigration and emigration rates can be calculated using Equations 6 and 7. Note that it is not the only method for calculating migration rates and there are other different options to assign them based on different species models (16).

$$\lambda_i = I\left(1 - \frac{C_i}{N}\right) \tag{6}$$

$$\mu_i = \left(\frac{E \times C_i}{N}\right) \tag{7}$$

where $\lambda_i$= immigration rate of habitat $i$, $\mu_i$= immigration rate of habitat $i$, $I$ = maximum immigration rate, $E$ = maximum emigration, $C_i$= species count of habitat $i$, $N$= maximum possible species count in any habitat.

If $\boldsymbol{SIV_i}$ is an $S$-dimensional integer vector of SIVs of habitat $i$ and there are $H$ habitats in the ecosystem, migration of SIVs among habitats can be implemented according to the Pseudo Code 1. In Pseudo Code 1, rand(0,1) is a uniformly distributed random real number in (0, 1).

*Pseudo Code 1 (migration)*

---

For i=1 to *H*
  Select $\mathbf{SIV_i}$ with probability $\propto \lambda_i$
  If rand (0, 1) $< \lambda_i$ then

    For j=1 to *H* do
      Select $\mathbf{SIV_j}$ with probability $\propto \mu_j$
      If rand (0, 1) $< \mu_j$ then
        Randomly select an SIV $\sigma$ from $\mathbf{SIV_j}$
        Replace a random SIV in $\mathbf{SIV_i}$ with $\sigma$
      End if
    End for
  End if
End for

---

### 3.1.2 Mutation

Mutation is performed to simulate sudden events like catastrophes. The mutation rate is calculated using Equation 8.

$$m_i = M\left(1 - \frac{P_i}{P_{max}}\right) \tag{8}$$

where $m_i$ = mutation rate of habitat *i*, $M$= maximum mutation rate, $P_i$= species count probability of habitat *i*, and $P_{max}$= maximum of all $P_i$. To know more about mutation and $P_i$ calculation, reader can refer to original paper by Simon (16). The mutation process is described in Pseudo Code 2. In this code, $SIV_i(j)$ is the *j*th SIV of the habitat *i*.

*Pseudo Code 2 (Mutation)*

---

For i=1 to *H*
  Compute the probability $P_i$
  Select $SIV_i(j)$ with probability $\propto P_i$
  If rand(0, 1) $< m_i$ then
    Replace $SIV_i(j)$ with a randomly generated SIV
  End if
End for

---

### 3.1.3 Elitism

The concept of elitism suggests that after each generation, habitats are sorted according to their HSI and top $Q$ (elitism parameter) habitats with highest HSI are taken to the next generation. This operation ensures that better habitats stay unspoiled in the subsequent generations.

### 3.1.4 Algorithm

Combining all components together, the BBO algorithm is presented in Figure 4. The pre-specified parameters are taken as input to BBO algorithm. In the initialization step, $S$ randomly selected SIVs (feature/variables) are assigned to each of the $H$ habitats. Subsequently, HSIs of all habitats are computed using Equation 5. In algorithmic flow, HSI is available first and therefore, there is a need to define a function to obtain number of species ($C_i$) from HSI. Equation 9 defines this mapping according to the HSI-based rank of a habitat ($RANK_i$), suggesting a plausible behavior that habitat with high HSI would have more number of species.

$$C_i = \{N - RANK_i | N \geq S\} \qquad (9)$$

After the calculation of species counts, emigration rate, immigration rate and species count probability are calculated to execute migration and mutation on non-elite habitats (as per Pseudo Codes 1 and 2). Subsequently, HSI of the modified habitats are recomputed using Equation 5 and then elitism is implemented. This procedure is performed till a pre-specified number of generations. At the end of simulation, SIVs corresponding to the habitat having the highest HSI are the important features. Section 4.2 systematically describes the integration of SVR and required modifications in this algorithm to make it useful for feature selection in the prediction models.

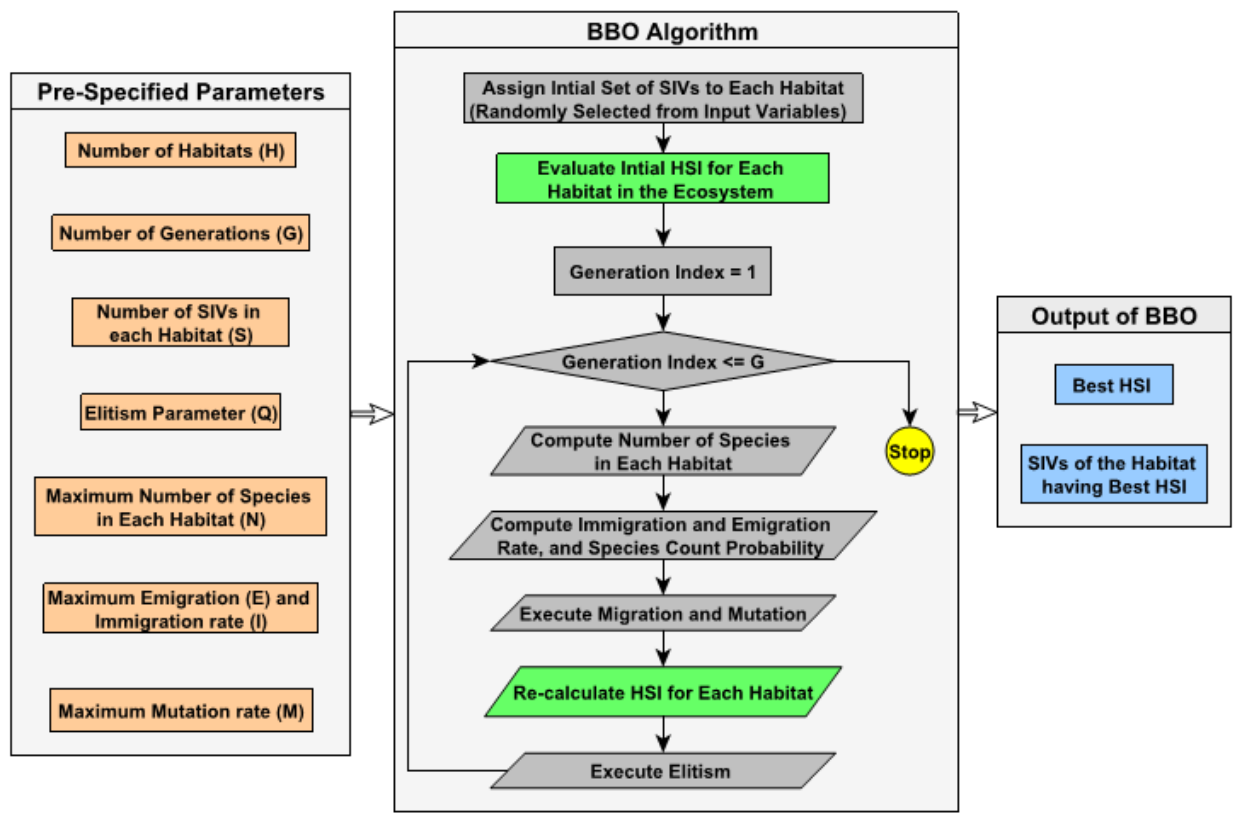

**FIGURE 4 Biogeography-based Optimization Algorithm**
Note: In the hybrid BBO-SVR approach, the green blocks in BBO algorithm are modified (Section 4.2).

## 3.2 Support Vector Regression (SVR)

Support Vector Machine (SVM) was developed to solve classification problems (20-21), but in current paradigm, it is extensively used to solve regression problems (22). There are two commonly used versions of SVR, 'epsilon-SVR' and 'nu-SVR'. The original SVR formulations use epsilon penalty parameter for points which are badly predicted. An alternative version of SVR was later developed where the epsilon penalty parameter was replaced by an alternative parameter, 'nu', which applies a slightly different penalty (22). The 'nu' parameter represents an upper bound on the fraction of training samples which are badly predicted and a lower bound on the fraction of samples which are support vectors. Though 'epsilon' and 'nu' are just different versions of the penalty parameter, this study uses 'nu-SVR' because it performs better and has a meaningful interpretation. Further description about the application of SVR in this study is given in the next section.

## 4 SVR AND BBO-SVR APPROACH FOR TRAVEL TIME PREDICTION

This section first illustrates the development of a travel time prediction model using stand-alone SVR approach and later demonstrates the feature selection and travel time prediction using a hybrid BBO-SVR approach.

### 4.1 Travel Time Prediction Model Using SVR

The data collected from National Taiwan Freeway No. 1 consist of 43 predictors (Table 1) and 7908 observations. As the number of features in input data was much less than observations, RBF kernel was adopted and optimal parameters of the kernel was found using grid-search approach (23). This study used the LIBSVM software suite (24) for implementing 'nu-SVR'.

First, observations were split in training and testing datasets. Out of the 7908 observations, 5270 (~66.6%) observations constituted to a training data set and the remaining 2638 (~33.4%) observations were used to test the trained model. Subsequently, both training and testing data were scaled between -1 and 1. Then model was trained using the training dataset with all 43 variables and the trained model was used to predict the travel time for the testing dataset. Then mean absolute percentage error (MAPE) of prediction for the testing dataset was calculated from Equation 10. The resultant MAPE of SVR prediction model is shown in Table 3 (highlighted in orange).

$$MAPE = \frac{1}{M} \sum_{1}^{M} \left| \frac{x(k) - y(k)}{x(k)} \right| \times 100\% \qquad (10)$$

Where $M$ is number of observations in the sample, $x(k)$ is the actual value, and $y(k)$ is the predicted value.

### 4.2 Proposed Hybrid BBO-SVR Approach for Prediction and Feature Selection

The objective of the hybrid BBO-SVR approach is to identify the set of important predictors for developing a travel time prediction model having prediction accuracy equivalent to that of stand-alone SVR prediction model (using all predictors). The predictors (flow, speed, rain fall and HTT) and MAPE are analogous to the SIVs and HSI in BBO terminology (Section 3.1), respectively. In the proposed hybrid approach, SVR is integrated with BBO algorithm while calculating HSI (or MAPE) of a habitat using Equation 5 (green blocks in Figure 4) and the rest of the BBO algorithm, shown in the Figure 4, remains intact.

The number of SIVs in each habitat ('S' in Figure 4), i.e., the desired number of important features (predictor) is specified before starting BBO-SVR simulation. In other words, a planner instructs the BBO-SVR framework as follows- "I want to predict travel time with, say five predictors and therefore, give me five best predictors (SIVs) and the corresponding MAPE (HSI) of prediction." After simulating pre-specified number of generations ('G' in Figure 4) of the BBO-SVR algorithm, the predictor variables (SIVs) corresponding to the habitat having lowest MAPE (highest HSI) are the important features. Thus, important predictors and prediction accuracy of model (MAPE) were simultaneously obtained while predicting travel time with these important predictors.

## 5 NUMERICAL EXPERIMENTS ON THE STUDY AREA APPLICATION

### 5.1 Numerical Experiments
The proposed approach was implemented in MATLAB R2009a. The parameter settings are described in Table 2.

**TABLE 2 Parameter Settings in BBO-SVR algorithm**

| Pre-specified Parameters | Value |
|---|---|
| Number of Habitats ($H$) | 50 |
| Number of Generations ($G$) | 20 |
| Number of SIVs in each Habitat ($S$) | Varied from 1 to 10 |
| Elitism Parameter ($Q$) | 2 |
| Maximum Number of Species in Habitat ($N$) | Equal to $S$ (Assumption) |
| Maximum Emigration Rate ($E$) | 1 |
| Maximum immigration Rate ($I$) | 1 |

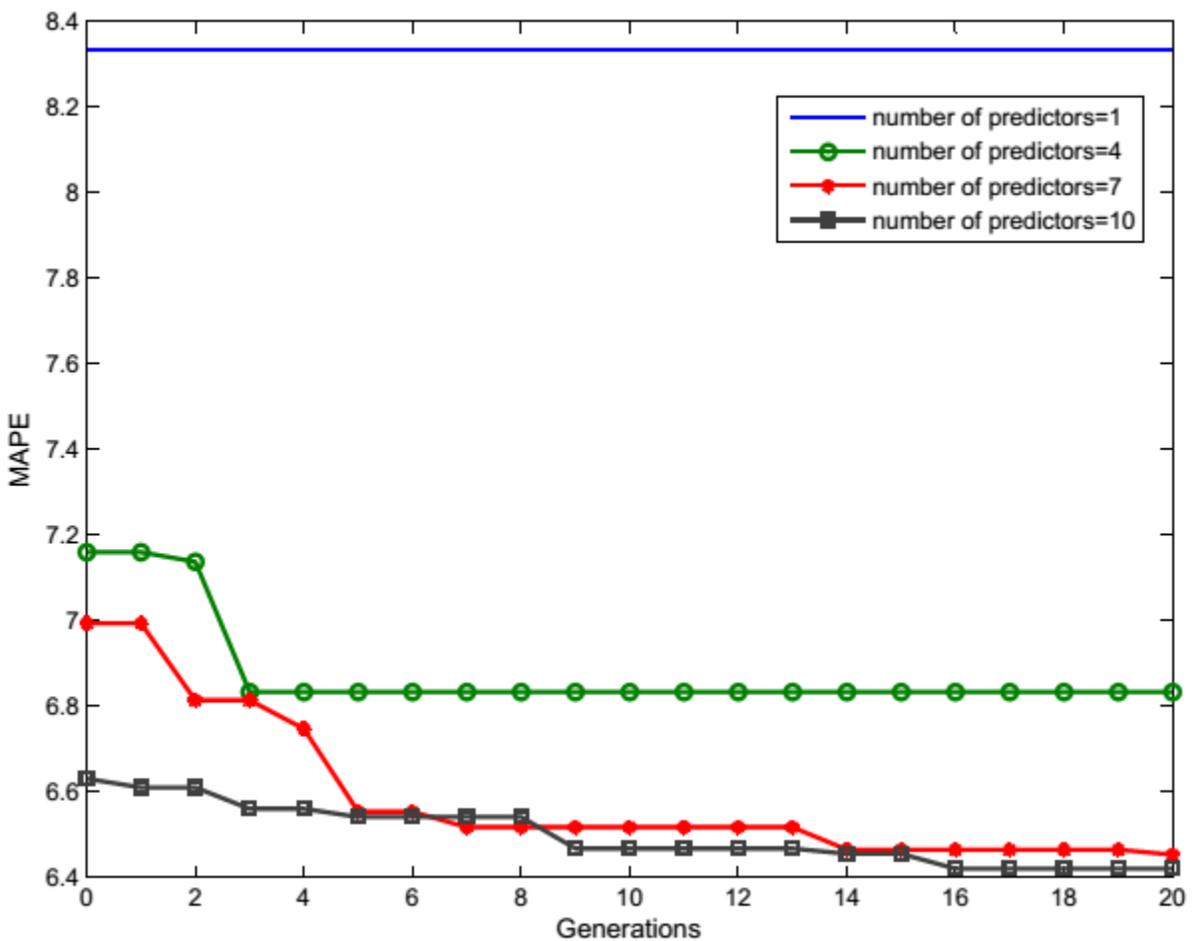

**FIGURE 5 Variation of best MAPE with progress of generations**

At first, One does not know that how many predictors (*S*) would be adequate to predict travel time using the hybrid BBO-SVR approach with an accuracy equivalent to that of the stand-alone SVR prediction model (using all forty three predictors). Therefore, 10 scenarios with different number of predictors, varying from 1 to 10, were considered, i.e., the BBO-SVR approach was instructed to give 1 best predictor & the corresponding MAPE in the first scenario, a set of 2 best predictors & the corresponding MAPE in the second scenario, and so forth. For each scenario, this study considered an ecosystem having 50 habitats (50 sets of solutions) and simulated the evolution of the ecosystem till 20 generations. While simulating the evolution of ecosystem, 2 habitats (elitism parameter, *Q=2*) having best MAPEs in the present generation, were not passed through migration and mutation operations to the subsequent generation.

After every generation of each scenario, the most suitable habitat, i.e. the one with lowest MAPE (highest HSI) was tracked to ensure the convergence of the algorithm. Figure 5 shows that the MAPE of the most suitable habitat is decreasing with generation, reflecting convergence of BBO-SVR algorithm. The convergence is visible in all scenarios, but in Figure 5, only four specific scenarios are included to maintain visual aesthetics. In all scenarios, decrease in the best MAPE with progress of generations becomes insignificant after the 10th generation, justifying that simulating the evolution of ecosystem up to 20 generations is enough for this problem instance.

To validate the feature selection utility of the proposed approach, travel time on the selected freeway section was also predicted by stand-alone SVR model, as described in Section 4.1. The next subsection compares the results of all 10 scenarios of the BBO-SVR approach with one another as well as with the results of the stand-alone SVR approach.

## 5.2 Results

Table 3 shows the MAPE of the stand-alone SVR model as well as the MAPE and the corresponding important predictors in all 10 scenarios of the BBO-SVR approach. The stand-alone SVR model predicts travel time with all forty three predictors with a MAPE of 6.46 (highlighted in Orange), but almost the same accuracy (MAPE of 6.44) can be achieved using only six predictors identified by the hybrid BBO-SVR approach (highlighted in Yellow). This outcome supports and validates the feature selection functionality of the proposed approach.

Surprisingly, even one predictor variable, i.e., HTT is able to predict freeway travel time with a MAPE of 8.33 (highlighted in Green), which is approximately 2 percent higher than the best prediction MAPE. Planners may thus consider using HTT as the only predictor for future travel time prediction. However, they should be well aware of the fact that the 2 percent higher MAPE, which may appear small in magnitude, may create a significant incredulity about information system among travelers.

If all predictors are independent then one may expect that the important predictors in one scenario should remain important in all the other scenarios. This pattern appears for a few predictors, for e.g. HTT was found to be the best predictor in all scenarios. However, such patterns are not consistent across all predictors, suggesting a correlation among them. Such correlations are expected in the speed and flow related variables collected from neighboring VDs.

It is worthy to note that only HTT and speed related variables play an important role in the prediction accuracy of the model whereas, rainfall and flow related variables are not important predictors because inclusion of them does not affect the MAPE of prediction significantly. The MAPE was expected to decrease with the inclusion of more predictors which is consistent in the first 6 scenarios. However, in later scenarios, inclusion of additional predictors was found to cause random insignificant variation (increase or decrease) in the MAPE values. These slight variations in MAPE are perhaps not caused due to characteristics of predictor variables but may be attributed to the random initialization and implicit randomness in migration and mutation steps of the algorithm.

## 6 CONCLUSIONS

This study developed a SVR-BBO hybrid approach for feature selection and long-distance freeway travel time prediction under non-recurrent congestion. The integrated framework was validated by predicting travel time on a 36.1 km long segment of National Taiwan Freeway No. 1. The proposed hybrid approach was found to be superior over a stand-alone SVR approach as it can predict travel time with almost the same accuracy but with only six predictors instead of forty three predictors. The results of the study area application indicate that the collection of historical travel time and speed variables is crucial for the development and installation of future advanced travel time information systems. It can be concluded that the effects of rainfall and flow variables are captured by historical travel time and speed variables and therefore, they can be excluded from future data collections. The collection of only important variables can potentially lead to significant equipment and monetary savings during future data collection. The framework developed is transferrable to other locations but care should be taken during variable selection because it is a function of several characteristics of the study area. The shock wave speed and queue length can improve the accuracy of travel time prediction in non-recurrent congestion conditions. These variables can be included in future research to investigate their predictive power.

**TABLE 3 Results of BBO-SVR and stand-alone SVR approach**

| Number of Predictors | MAPE | The important predictors (SIVs) | | | | | | | | | |
|---|---|---|---|---|---|---|---|---|---|---|---|
| **1** | **8.33** | **HTT** | | | | | | | | | |
| 2 | 7.42 | Speed6230 | HTT | | | | | | | | |
| 3 | 7.01 | Speed6800 | Speed6230 | HTT | | | | | | | |
| 4 | 6.83 | Speed6625 | Speed6230 | Speed5160 | HTT | | | | | | |
| 5 | 6.62 | Speed6800 | Speed6230 | Speed5940 | Speed5160 | HTT | | | | | |
| **6** | **6.44** | **Speed6800** | **Speed6625** | **Speed6230** | **Speed5160** | **Speed4530** | **HTT** | | | | |
| 7 | 6.49 | Speed6800 | Speed6625 | Speed6230 | Speed5940 | Speed5160 | Speed4920 | HTT | | | |
| 8 | 6.45 | Speed6800 | Speed6550 | Speed6230 | Speed6150 | Speed5160 | Flow5160 | Speed4530 | HTT | | |
| 9 | 6.50 | Speed6800 | Speed6550 | Speed6230 | Speed5940 | Speed5160 | Speed5080 | Speed4920 | Taoyuan_Raining_3HR | HTT | |
| 10 | 6.42 | Speed6800 | Speed6625 | Speed6230 | Flow6150 | Speed5160 | Speed4530 | Yangmei_Raining_2HR | Taoyuan_Raining_1HR | Taoyuan_Raining_6HR | HTT |
| **43** | **6.46** | **All forty-three variables** | | | | | | | | | |

*Note:* In the variable 'Speed6800', 6800 represents the detector number, 'Taoyuan _Raining_2HR': cumulative rainfall of past 2 hours at Taoyuan station, 'HTT': Historical Travel time.

**ACKNOWLEDGEMENTS**
This work is partially supported by Dr. Mu-Chen Chen (Professor at National Chiao Tung University) and National Science Council, Taiwan, R.O.C. under grant NSC 100-2410-H-009-013-MY3. I am also thankful to Dr. Chen for providing data for calibrating the proposed method.